\documentclass[aps,showpacs, prl,twocolumn,superscriptaddress]{revtex4}%
\usepackage{epsfig,dsfont,amssymb,amsmath,amsthm,amsfonts,amsbsy,mathrsfs}
\usepackage{graphicx}
\usepackage{amsmath}
\usepackage{amssymb}
\begin{document}
\title{First-principles study of a novel superhard boron nitride phase}
\author{Chaoyu He}
\affiliation{Laboratory for Quantum Engineering and Micro-Nano
Energy Technology, Xiangtan University, Xiangtan 411105, China}
\author{L. Z. Sun}
\email{lzsun@xtu.edu.cn} \affiliation{Laboratory for Quantum
Engineering and Micro-Nano Energy Technology, Xiangtan University,
Xiangtan 411105, China}
\author{C. X. Zhang}
\affiliation{Laboratory for Quantum Engineering and Micro-Nano
Energy Technology, Xiangtan University, Xiangtan 411105, China}
\author{Xiangyang Peng}
\affiliation{Laboratory for Quantum Engineering and Micro-Nano
Energy Technology, Xiangtan University, Xiangtan 411105, China}
\author{K. W. Zhang}
\affiliation{Laboratory for Quantum Engineering and Micro-Nano
Energy Technology, Xiangtan University, Xiangtan 411105, China}
\author{Jianxin Zhong}
\email{jxzhong@xtu.edu.cn}\affiliation{Laboratory for Quantum
Engineering and Micro-Nano Energy Technology, Xiangtan University,
Xiangtan 411105, China}
\date{\today}

\begin{abstract}
A superhard boron nitride phase dubbed as Z-BN is proposed as
possible intermediate phase between h-BN and zinc blende BN (c-BN),
and investigated using first-principles calculations within the
framework of the density functional theory. Although the structure
of Z-BN is similar to that of bct-BN containing four-eight BN rings,
it is more energy favorable than bct-BN. Our study reveals that
Z-BN, with a considerable structural stability and high density
comparable to c-BN, is a transparent insulator with an indirect band
gap about 5.27 eV. Amazingly, its Vickers hardness is 55.88 Gpa
which is comparable to that of c-BN. This new BN phase may be
produced in experiments through cold compressing AB stacking h-BN
due to its low transition pressure
point of 3.3 GPa.\\
\end{abstract}
\maketitle \indent Recently, some superhard allotropes of carbon,
such as the monoclinic M-carbon \cite{M3, 16}, cubic body center C4
carbon (bct-C4) \cite{171, 172, 17}, orthorhombic W-carbon \cite{18}
and orthorhombic Z-carbon \cite{19, 20, 21}, were proposed as the
candidates for the superhard phase observed in cold compressing
graphite \cite{15}. These studies are significant for understanding
the transformation between graphite and diamond and arouse many
interests in searching for low energy superhard carbon phase
\cite{n1, n2, n3, n4, n5, n6, n7, n8, n9}. Lots of theoretical
efforts had been paid on searching for low energy superhard carbon
allotropes, such as the particle-swarm optimization method \cite{M1,
MM1}, graph theoretical methods \cite{M2}, evolutionary algorithm
USPEX \cite{M3}, minima hopping method \cite{M4} (MHM) for crystal
structure prediction \cite{M5} and fragment assembly method used by
Niu \cite{n4}, Zhou \cite{n5} and in our latest work \cite{n8}.
Interestingly, the superhard bct-C4 \cite{171} and M-carbon
\cite{M1} were predicted before the cold compressing graphite
experiment and later they are identified as the candidates \cite{16,
17} for the superhard graphite.\\
\indent In comparison with the efforts paid on new carbon phases,
there are relative less theoretical attentions on the research for
superhard BN phases. Actually, the polymorphs of boron nitride, such
as hexagonal BN (h-BN) \cite{1}, zinc blende BN (c-BN)\cite{2},
wurtzite BN (w-BN) \cite{3}, BN nanotubes \cite{4}, BN fullerence
\cite{5} and amorphous BN \cite{6}, can be regarded as the
counterpart systems of graphite, cubic-diamond (C-diamond),
hexagonal-diamond (H-diamond), carbon nanotubes (CNTs), fullerence
and amorphous. The superhard w-BN and c-BN have been arising intense
interests owing to their excellent optical, electrical and
mechanical properties \cite{7}. The phase transition from
compressing the soft h-BN to superhard c-BN and w-BN has been
interesting issue for decades and attracted many theoretical
\cite{8, 9, 10, 11, 111} and experimental efforts \cite{4, 12, 13}.
To date, the mechanism of such transitions is still ambiguous. For
example, through cold compressing h-BN the product is always the
meta-stable w-BN instead of the stable c-BN \cite{12}. Wen et al.
\cite{14} suggested that there might be some intermediate states
between h-BN and w-BN, and proposed a new BN phase (bct-BN)
\cite{14, bctBN2, bctBN3} with considerable stability and excellent
mechanical properties. Similar to the cold compression of graphite,
cold compressing h-BN may results in some unknown intermediate
states. More theoretical attentions are needed to search for the
potential intermediate
BN phases in cold compressing h-BN.\\
\indent In view of the similar characteristics between carbon and BN
systems, all the previously proposed carbon allotropes might be
excellent templates for finding new BN phases. It is worthy to
investigate the stability, mechanical and electronic properties of
new BN phases with the structures as M-carbon, bct-C4, W-carbon,
Z-carbon and other proposed carbon structures due to the
comparability between graphite and h-BN. The transforming path from
h-BN to bct-BN has been studied by Wen et al. \cite{14}. According
to their results, M-BN is energy unstable due to the existence of
the boron-boron (B-B) and nitrogen-nitrogen (N-N) bonds\cite{14}. BN
phases with the structure of M-carbon, W-carbon, S-carbon and
H-carbon are expected to be unstable due to the existence of
five-seven patterns in such structures. Similar to bct-C4, the
recently theoretically proposed structure Z-carbon \cite{19, 20, 21}
is an excellent template for new BN phase because its structure
contains only even carbon rings (four-eight carbon rings). In our
present work, based on the first-principle calculations, we
systematically investigate the stability, electronic and mechanical
properties of the BN allotrope with the structure of Z-carbon (named
as Z-BN). The result is an interesting example of extending the
previously predicted carbon allotropes to their corresponding BN
counterpart. Such method is an effective approach to search for new
superhard BN phases.\\
\section{Computational Details}
\indent All calculations are carried out using the density
functional theory with both local density approximation (LDA)
\cite{lda,lda1} and general gradient approximation (GGA) \cite{gga}
as implemented in Vienna ab initio simulation package (VASP)
\cite{22, 23}. The interactions between nucleus and the valence
electrons of boron and nitrogen are described by the projector
augmented wave (PAW) method \cite{24, 25}. A plane-wave basis with a
cutoff energy of 500 eV is used to expand the wave functions of all
systems considered in our present work. The Brillouin Zone (BZ)
sample meshes for all systems are set to be denser enough
(11$\times$11$\times$9 for h-BN, 9$\times$9$\times$9 for C-BN,
13$\times$13$\times$7 for w-BN, 7$\times$7$\times$13 for bct-BN and
5$\times$9$\times$11 for Z-BN) in our calculations. Crystal lattices
and atoms positions of h-BN, w-BN, c-BN, bct-BN, and Z-BN are fully
optimized (under different external pressure) up to the residual
force on every atom less than 0.005 eV/{\AA} through the
conjugate-gradient algorithm. Vibration properties of all systems
are studied by using the phonon package \cite{26} with the forces calculated from VASP.\\
\begin{figure}
\includegraphics[width=3.5in]{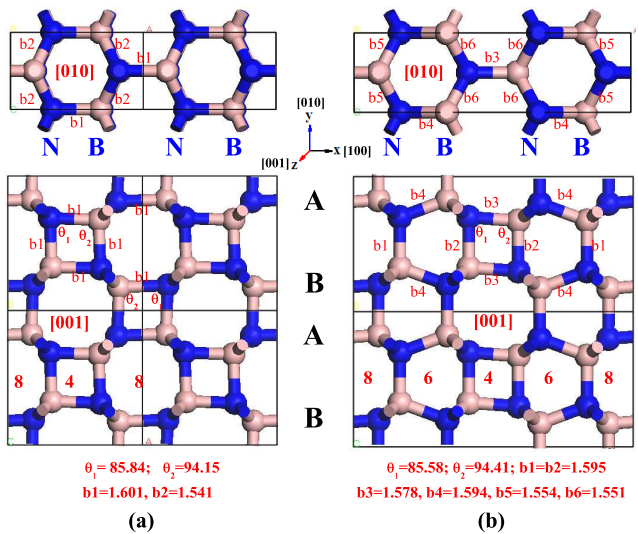}\\
\caption{ Views from [010] direction (top) and [001] direction
(bottom) of bct-BN (a) and Z-BN (b).}\label{fig1}
\end{figure}
\indent To evaluate the transition pressure from h-BN to Z-BN, the
exchange-correlation functional is describe by LDA. Although the LDA
is a simple approximation of DFT, it can give reasonable interlayer
distance, mechanical properties of h-BN sheets due to a delicate
error cancelation between exchange and correlation in comparison
with that of semi-local generalized gradient approximation (GGA).
The detail comparison of the lattice parameters of h-BN derived from
GGA and LDA
can be found in our previous report \cite{hu}.\\
\indent Benchmark calculations are conducted for c-BN phase to
validate our computational scheme. The calculated lattice parameter
with GGA is 3.625 {\AA}, the result agrees well with the
experimental value of 3.620 {\AA} \cite{3}. The computed elastic
constants for the c-BN phase with GGA are C$_{11}$ = 780 GPa,
C$_{44}$ = 444 GPa, and C$_{12}$ = 173 GPa. They are also in
reasonable agreement with the experimental values of C$_{11}$= 820
GPa, C$_{44}$ = 480 GPa,
and C$_{12}$ = 190 GPa \cite{27}.\\
\begin{table*}
  \centering
  \caption{Space group, lattice information (LP), density (D: g/cm$^{3}$), band gap (Eg: eV), cohesive energy (Ecoh: eV/BN), bulk modulus (B${_0}$: Gpa),
  shear modulus (G: Gpa) and Vickers hardness (H${_v}$: Gpa) for the c-BN,  w-BN, bct-BN and Z-BN.}\label{tabI}
\begin{tabular}{c c c c c c c c c c}
\hline \hline
Systems &Space group &LP                                  &D        &Eg      &Ecoh     &B${_0}$       &G         &H${_v}$  &reference\\
\hline
c-BN   &F-43m   &a=b=c=3.625{\AA}                          &3.593   &4.40    &-6.934   &376.19   & 381.52   &62.82     &this work    \\
       &        &a=b=c=3.620{\AA}                          &        &        &         &         &          &          &experimental \cite{3}  \\
       &        &a=b=c=3.589{\AA}                          &        &        &         &         &          &          &calculated \cite{14}   \\
w-BN   &P63mc   &a=b=2.555{\AA}, c=4.225{\AA}              &3.587   &5.24    &-6.930   &375.24   & 384.17   &63.82     &this work    \\
       &        &a=b=2.550{\AA}, c=4.200{\AA}                &        &        &         &         &          &          &experimental \cite{3} \\
       &        &a=b=2.538{\AA}, c=4.179{\AA}              &        &        &         &         &          &          &calculated \cite{14}   \\
bct-BN &P42/mnm &a=b=4.425{\AA}, c=2.548{\AA}              &3.431   &4.83    &-6.845   &348.35   & 309.44   &46.86     &this work    \\
       &        &a=b=4.380{\AA}, c=2.526{\AA}              &        &        &         &         &          &          &calculated \cite{14}   \\
Z-BN   &Pbam    &a=8.891{\AA}, b=4.293{\AA}, c=2.555{\AA}  &3.520   &5.27    &-6.872   &359.61   & 347.45   &55.88    &this work    \\
\hline \hline
\end{tabular}
\end{table*}
\begin{figure}
\includegraphics[width=3.5in]{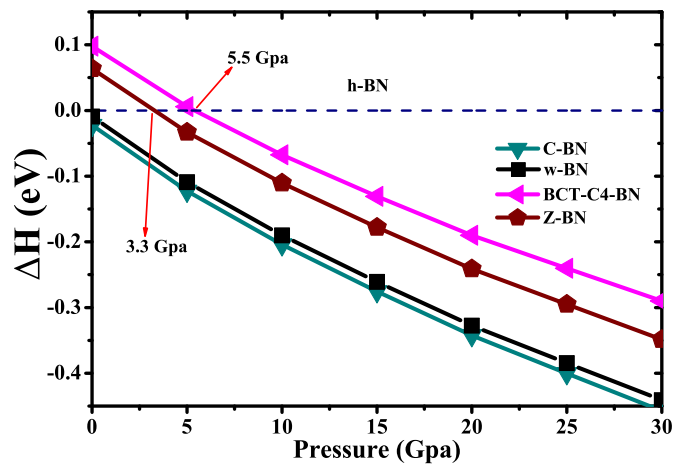}\\
\caption{The enthalpy per BN for c-BN, w-BN, bct-BN and Z-BN as a
function of pressure relative to h-BN (Derived from LDA
calculations).}\label{fig2}
\end{figure}
\section{Results and Discussions}
\indent The crystal structures of bct-BN and Z-BN are shown in
Fig.~\ref{fig1}. The lattice constants of c-BN, w-BN, bct-BN and
Z-BN derived from GGA at zero pressure are listed in
Tab.~\ref{tabI}. Fig.~\ref{fig1}(a) shows the [010] (top) and [001]
(bottom) direction views of bct-BN. The crystal structure of bct-BN
belongs to P42/mnm space group. At zero pressure, its equilibrium
lattice constants derived from GGA are a=b=4.425 {\AA} and c=2.548
{\AA}. One inequivalent B atom in its unit cell occupies the Wyckoff
position at 4g (0.325, 0.3675, 0.0) and one inequivalent N atom
locates at 4f (0.313, 0.313, 0.0). There are two inequivalent B-N
bonds labeled as b$_1$ and b$_2$ with length of 1.601 {\AA} and
1.541 {\AA} in bct-BN. Their average length is 1.571 {\AA}. Z-BN
belongs to Pbam space group and its equilibrium lattice constants
are a=8.891 {\AA}, b=4.293 {\AA} and c=2.555 {\AA}. Two inequivalent
N atoms in Z-BN occupy the Wyckoff positions at 4g (0.834, 0.301,
0.0) and 4h (0.589, 0.301, 0.5). Another two inequivalent B atoms in
Z-BN occupy 4g (0.334, 0.827, 0.0) and 4h (0.088, 0.827, 0.5)
Wyckoff positions. The views from [010] (top) and [001] (bottom)
direction of Z-BN are shown in Fig.~\ref{fig1}(b). There are six
inequivalent B-N bonds in Z-BN, labeled as b$_1$, b$_2$, b$_3$,
b$_4$, b$_5$ and b$_6$, as shown in Fig.~\ref{fig1}(b). Their bond
length are 1.595 {\AA},  1.595 {\AA}, 1.578 {\AA},  1.594 {\AA},
1.554 {\AA} and 1.551 {\AA}, respectively. The average length is
1.578 {\AA}. The average bond lengths of both bct-BN and Z-BN are
comparable to that of diamond (1.570 {\AA}). We define the bond
angles BNB and NBN in the parallelogram of both systems as
${\theta}$${_1}$ and ${\theta}$${_2}$, respectively. These two
angles are ${\theta}$${_1}$=85.84 and ${\theta}$${_2}$=94.16 for
bct-BN
and ${\theta}$${_1}$=85.58 and ${\theta}$${_2}$=94.42 for Z-BN.\\
\indent The apparent difference between these two new BN phases is
the absence of hexagon pattern for bct-BN viewed from their [001]
direction. The structures of both bct-BN and Z-BN are constructed
with the four-eight patterns and without the five-seven patterns as
those in M-carbon and W-carbon. From the structural point of view,
both bct-BN and Z-BN can be regarded as the mutations of w-BN, as
their corresponding counterpart systems bct-C4 and Z-carbon are the
mutations of H-diamond \cite{n8}. Moreover, both bct-BN and Z-BN can
be derived from reconstructing the AB stacking h-BN with different
manners, as the potential
intermediate products in cold compressing AB stacking h-BN.\\
\begin{figure}
\includegraphics[width=3.50in]{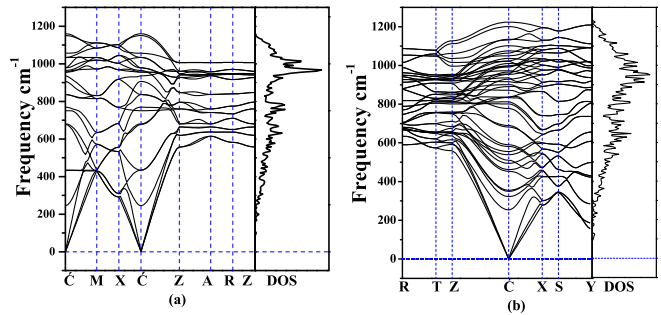}
\caption{Phonon band structure and phonon density of states of
bct-BN (a) and Z-BN (b) at Zero pressure.}\label{fig3}
\end{figure}
\indent The relative stability of c-BN, w-BN, bct-BN and Z-BN is evaluated through comparing their cohesive energy per BN pair. Bct-BN has been predicated more favorable than M-type BN phase\cite{14}. According to our GGA results, we find that Z-BN (-6.872 eV/BN) is more favorable than bct-BN (-6.845 eV/BN), and its cohesive energy is only 62 meV per BN above c-BN (-6.934 eV/BN). The enthalpy per BN pair for c-BN, w-BN, bct-BN as well as Z-BN as functions of pressure relative to AB stacking h-BN derived from LDA calculation is shown in Fig.~\ref{fig2}. The results indicate that when the pressure is larger than 3.3 GPa (5.5 Gpa), Z-BN (bct-BN) is more stable than h-BN. Namely, the transition pressure point of Z-BN (bct-BN) from h-BN under external pressure is 3.3 GPa (5.5 Gpa). Moreover, Z-BN is always more favorable than bct-BN as the external pressure ranging from 0 to 30 GPa. To further confirm the dynamic stability of bct-BN and Z-BN, we calculated their phonon band structures and phonon density of states. The results derived form GGA are shown in Fig.~\ref{fig3} (a) and (b) for bct-BN and Z-BN, respectively. For both systems, there is no negative frequency and states in phonon band structure and phonon density of states, confirming the dynamic stability of bct-BN and Z-BN.\\
\indent The space group, density, band gap, cohesive energy, bulk modulus, shear modulus and Vicker's hardness of c-BN, w-BN, bct-BN and Z-BN are summarized in Tab.~\ref{tabI}. The results of density reveal that Z-BN (3.520 g/cm$^{3}$) is denser than bct-BN ( 3.431 g/cm$^{3}$ ), and its density is comparable to those of w-BN (3.587 g/cm$^{3}$) and c-BN (3.593 g/cm$^{3}$). The values of bulk modulus of bct-BN (348.35 GPa) and Z-BN (359.61 Gpa ) are comparable to those of c-BN (376.19 Gpa) and w-BN (375.24 Gpa). To further analyze the hardness of Z-BN, we adopt the recently introduced empirical scheme \cite{28} to evaluate the Vickers hardness (H${_v}$) determined by the bulk modulus (B${_0}$) and shear modulus (G), where: H${_v}$=2(G${^3}$/B${^2_0}$)${^{0.585}}$-3. The values of Vickers hardness for bct-BN, Z-BN, w-BN and c-BN are 46.86 Gpa, 55.88 GPa , 63.82 Gpa and 62.82 Gpa, respectively. The results indicate that Z-BN is a superhard material comparable to c-BN.\\
\begin{figure}
\includegraphics[width=3.50in]{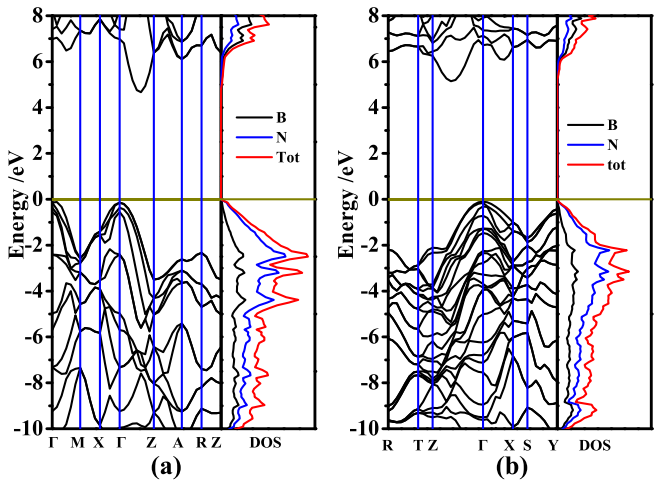}\\
\caption{Electronic band structures and density of states of bct-BN
(a) and Z-BN (b) at zero pressure obtained by GGA.}\label{fig4}
\end{figure}
\indent To investigate the electronic properties of bct-BN and Z-BN,
their band structures and density of states are calculated and shown
in Fig.~\ref{fig3}(a) and (b), respectively. The results indicate
that the valence band maximum (VBM) and conduction band minimum
(CBM) of Z-BN (bct-BN) are mainly derived from the N atoms and B
atoms, respectively. Moreover, both systems are
indirect-wide-band-gap insulators. The band gap of bct-BN is 4.83 eV
which is in good agreement with previous first-principles
calculation \cite{14}. The band gap of Z-BN is 5.27 eV
which is bigger than those of c-BN, w-BN and bct-BN. Namely, both Z-BN and bct-BN are transparent superhard materials.\\
\begin{figure}
\includegraphics[width=3.50in]{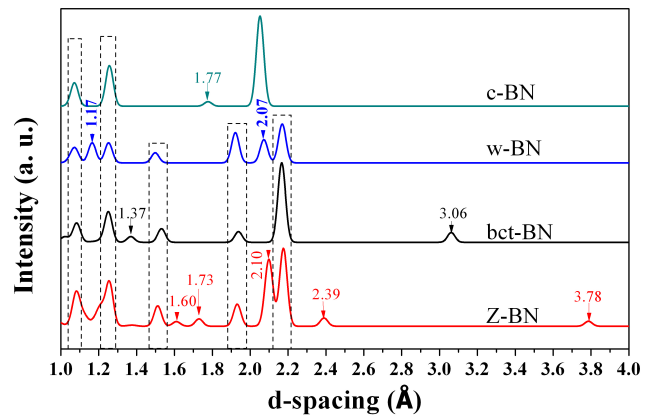}\\
\caption{The simulated X-ray diffraction (XRD) patterns for C-BN,
w-BN, bct-BN and Z-BN at 10 Gpa.}\label{fig4}
\end{figure}
\indent Finally, we provide the simulated X-ray diffraction (XRD)
patterns for C-BN, w-BN, bct-BN and Z-BN at 10 Gpa (above all the
phase transition pressures) to identify their potential products.
Each phase have their characteristic peaks. C-BN has a
characteristic peak located at 1.77 {\AA}. w-BN holds characteristic
peaks at d-spacing of 1.17 {\AA} and 2.07 {\AA}, both of them are
absent in bct-BN and Z-BN. There are two characteristic peaks for
bct-BN located at d-spacing of 1.37 {\AA} and 3.06 {\AA},
respectively. Five characteristic peaks of Z-BN are located at
d-spacing of 1.60 {\AA}, 1.73 {\AA}, 2.10 {\AA}, 2.39 {\AA} and 3.78
{\AA}. These peaks are absent in the other three BN phases. One can
identify Z-BN from other three phases according to these
characteristic peaks.\\
\section{Conclusion}
\indent The structural, vibrational, mechanical and electronic
properties of a new BN allotrope (Z-BN) with the structure of
Z-carbon has been systematically investigated using first-principles
calculations. Our results indicate that Z-BN is dynamically stable.
The value of hardness of Z-BN is larger than that of bct-BN and
comparable with those of w-BN and c-BN. Z-BN is a transparent
insulator with an indirect band gap about 5.27 eV. Under external
pressure up to 3.3 Gpa, Z-BN is energy more stable than h-BN. Our
results indicate that Z-BN
phase with remarkable stability may be the potential intermediate product in cold compressing AB stacking h-BN.\\
\section{Acknowledgements}
This work is supported by the National Natural Science Foundation of
China (Grant Nos. 10874143 and 10774127), the Cultivation Fund of
the Key Scientific and Technical Innovation Project, the Ministry of
Education of China (Grant No. 708068), the Program for New Century
Excellent Talents in University (Grant No. NCET-10-0169), the
Scientific Research Fund of Hunan Provincial Education Department
(Grant No. 10K065) and the Hunan Provincial Innovation Foundation
for Postgraduate (Grant No. CX2010B250).

\end{document}